\documentstyle[prb,aps,multicol,epsf]{revtex}

\begin{document}

\draft
\title{
Local perturbation in a Tomonaga-Luttinger liquid at $g=1/2$:\\
orthogonality catastrophe, Fermi-edge singularity, and local density
of states
}
\author{A. Furusaki}
\address{Yukawa Institute for Theoretical Physics, Kyoto University,
Kyoto 606-01, Japan}
\date{\today}
\maketitle
\begin{abstract}
The orthogonality catastrophe in a Tomonaga-Luttinger liquid with an
impurity is reexamined for the case when the interaction parameter or
the dimensionless conductance is $g=1/2$.
By transforming bosons back to fermions, the Hamiltonian is reduced to 
a quadratic form, which allows for explicit calculation of the overlap 
integral and the local density of states at the defect site.
The exponent of the orthogonality catastrophe due to a backward
scattering center is found to be 1/8, in agreement with previous
studies using different approaches.
The time-dependence of the core-hole Green's function is computed
numerically, which shows a clear crossover from a non-universal
short-time behavior to a universal long-time behavior.
The local density of states vanishes linearly in the
low-energy limit at $g=1/2$.
\end{abstract}

\begin{multicols}{2}

\section{Introduction}
\label{sec:intro}

One-dimensional interacting fermion systems, Tomonaga-Luttinger (TL)
liquids,\cite{Tomonaga,Luttinger,Haldane} have recently attracted
much attention due to their anomalous response to local perturbations.
Recent extensive studies
\cite{Kane,Furusaki,Matveev,Moon,Fendley,Leung} on transport
properties of TL liquids with an impurity revealed that repulsively
interacting fermions have vanishing transmission probability through a
potential barrier in the low-energy limit.
This is because the interaction between fermions strongly enhances the
backward scattering at the barrier.
Thus, a single defect effectively cuts a TL liquid into two
disconnected ones at zero temperature.\cite{Kane}
This implies that the local density of states (LDOS) at the defect is
reduced for low energy, and according to Kane and
Fisher\cite{Kane} it shows a power-law energy dependence,
\begin{equation}
\rho(\omega)\propto\omega^{\frac{1}{g}-1},
\label{rho(omega)}
\end{equation}
where $g$ is a parameter characterizing the TL liquid.
This picture was, however, questioned recently by Oreg
and Finkel'stein,\cite{Oreg1} who claimed based on a mapping to a
Coulomb gas problem that the LDOS at the defect is enhanced, rather
than suppressed, in the low-energy limit for weakly interacting fermions.
This controversy motivates us to reexamine this issue.

The orthogonality catastrophe\cite{Anderson} in a TL liquid is another 
interesting subject which has been discussed by several
authors.\cite{Ogawa,Lee,Gogolin,Prokofev,KMG,Affleck,Oreg2} 
They showed that the overlap between the ground state of a
pure TL liquid $|{\rm p}\rangle$ and that of a TL liquid with a single
scatterer $|{\rm s}\rangle$ vanishes in the limit of large system size:
\begin{equation}
|\langle{\rm p}|{\rm s}\rangle|^2\propto L^{-\gamma_F-\gamma_B},
\label{-gamma_F-gamma_B}
\end{equation}
where $L$ is the length of the system.
The exponent $\gamma_F$ is due to the forward-scattering potential and
depends on its strength.\cite{Ogawa,Lee}
It can be calculated directly using a unitary transformation.
The other exponent $\gamma_B$ due to the backward scattering is
believed to be independent of the strength of the potential and take a 
universal value, $1/8$.\cite{Gogolin,Prokofev,KMG,Affleck}
In Refs.~\onlinecite{Gogolin,Prokofev}, and \onlinecite{Affleck} the
exponent $\gamma_B$ is calculated by assuming that a backward scattering
center can be replaced with an impenetrable potential barrier.
Oreg and Finkel'stein,\cite{Oreg2} however, questioned the validity of 
the assumption and argued that the exponent of the Fermi-edge
singularity due to a backward scattering center is zero, which implies 
$\gamma_B=0$.
On the other hand, Kane {\it et al.}\cite{KMG} used a
renormalization-group equation which becomes exact in the limit of
weak repulsive interaction between fermions.
They could describe a crossover from the high-energy regime to the
low-energy regime, and obtained the same exponent $\gamma_B=1/8$ in
the low-energy limit.
The result of a recent direct numerical calculation of the overlap
integral\cite{Qin} is also consistent with $\gamma_B=1/8$.

It is known that, when the TL-liquid parameter $g$ is 1/2, the
bosonized Hamiltonian containing a nonlinear term representing the
backward scattering can be transformed to a quadratic Hamiltonian of
fermions.\cite{Matveev2}
This is essentially the same technique as the Emery-Kivelson solution
of the two-channel Kondo problem.\cite{Emery}
The exact results on the conductance\cite{Kane} and non-equilibrium
noise spectra\cite{Chamon} were obtained using this refermionization
technique.
It is thus natural to expect that exact calculation should also be
possible for the above-mentioned problems.
The purpose of this paper is to show that this is indeed the case.

The structure of this paper is as follows.
After introducing a model of interacting fermions in
Sec.~\ref{sec:model}, we discuss in Sec.~\ref{sec:ldos} the exact
low-energy behavior of the LDOS for $g=1/2$.
For $g\ne1/2$ we show that Eq.~(\ref{rho(omega)}) follows from the
assumption that the phase field is pinned at the defect site.
The importance of zero-modes is emphasized.
In Sec.~\ref{sec:ortho} we calculate $\gamma_B$ analytically for
$g=1/2$ without assuming the nature of the low-energy fixed point.
We find $\gamma_B=1/8$.
The so-called core-hole Green's function is then computed numerically
in Sec.~\ref{sec:core-hole}, which shows a clear crossover from
short-time to long-time regimes. 
We show in Sec.~\ref{sec:fermi-edge} that the exponent of the
Fermi-edge singularity due to backward scattering is also given by
$\gamma_B$. 
We summarize the results in Sec.~\ref{sec:conclusion}.

\section{Model}
\label{sec:model}
In this section we introduce a model of interacting spinless fermions
and briefly explain the bosonization rule to fix the notation.

The Hamiltonian of our model is given by
\end{multicols}
\vspace{-6mm}\noindent\underline{\hspace{87mm}}
\begin{eqnarray}
H&=&
iv_F\int^\infty_{-\infty}dx
\left[\psi^\dagger_L(x)\frac{d}{dx}\psi_L(x)
      -\psi^\dagger_R(x)\frac{d}{dx}\psi_R(x)\right]
+g_2\int^\infty_{-\infty}dx
:\!\psi^\dagger_L(x)\psi_L(x)\!:\, :\!\psi^\dagger_R(x)\psi_R(x)\!:\cr
&&
+\frac{g_4}{2}\int^\infty_{-\infty}dx\sum_{\mu=L,R}
  :\!\psi^\dagger_\mu(x)\psi_\mu(x)\!:
  :\!\psi^\dagger_\mu(x)\psi_\mu(x)\!:
+\lambda_F\sum_{\mu=L,R}:\!\psi^\dagger_\mu(0)\psi_\mu(0)\!:
+\lambda_B\!\left[e^{i\theta}\psi^\dagger_L(0)\psi_R(0)+{\rm h.c.}\right],
\label{H1}
\end{eqnarray}
\noindent\hspace{92mm}\underline{\hspace{87mm}}\vspace{-3mm}
\begin{multicols}{2}\noindent
where $\psi_{L(R)}$ describes left-going (right-going) fermions,
$:\!\!A\!\!:$ represents normal-ordered operator $A$, and $\lambda_F$
($\lambda_Be^{i\theta}$) is the forward-scattering
(backward-scattering) potential. 
Following the standard bosonization rule,\cite{Solyom} we express
fermions $\psi_\mu$ in terms of bosonic operators:
\begin{mathletters}
\begin{eqnarray}
\psi_L(x)&=&\frac{1}{\sqrt{2\pi\alpha}}\eta_Le^{-i\varphi_L(x)},\\
\psi_R(x)&=&\frac{1}{\sqrt{2\pi\alpha}}\eta_Re^{i\varphi_R(x)},\\
:\!\psi^\dagger_\mu(x)\psi_\mu(x)\!: \!&=&
\frac{1}{2\pi}\frac{d}{dx}\varphi_\mu(x),
\label{rule1}
\end{eqnarray}
\end{mathletters}\noindent
where $\alpha$ is a short-distance cutoff.
The bosonic fields satisfy the commutation relations
$[\varphi_L(x),\varphi_L(y)]=-i\pi{\rm sgn}(x-y)$,
$[\varphi_R(x),\varphi_R(y)]=i\pi{\rm sgn}(x-y)$, and
$[\varphi_L(x),\varphi_R(y)]=0$.
The operator $\eta_\mu$'s are Majorana fermions corresponding to zero
modes of bosons, which are needed to ensure the anticommutation
relation between $\psi_L$ and $\psi_R$.
They satisfy $\{\eta_L,\eta_R\}=0$ and $\eta^2_\mu=1$.
We then introduce new bosonic fields as
\begin{mathletters}
\begin{eqnarray}
\phi(x)&=&\frac{1}{\sqrt{4\pi}}[\varphi_R(x)+\varphi_L(x)],\\
\Pi(x)&=&-\frac{1}{\sqrt{4\pi}}\frac{d}{dx}[\varphi_R(x)-\varphi_L(x)],
\end{eqnarray}
\end{mathletters}\noindent
which obey $[\phi(x),\Pi(y)]=i\delta(x-y)$.
With these fields the Hamiltonian can be transformed to a bosonic form,
\begin{eqnarray}
H&=&
\frac{v}{2}\int dx
\left[\frac{1}{g}\left(\frac{d\phi}{dx}\right)^2
      +g\Pi^2\right]\cr
&&
+\frac{\lambda_F}{\sqrt{\pi}}\frac{d\phi(0)}{dx}
+i\frac{\lambda_B}{\pi\alpha}\eta_L\eta_R
  \sin\left[\sqrt{4\pi}\phi(0)+\theta\right].
\label{H2}
\end{eqnarray}
The parameter $g$ is related to $g_2$ and $g_4$ by
$g=[(1+\tilde g_4-\tilde g_2)/(1+\tilde g_4+\tilde g_2)]^{1/2}$ with
$\tilde g_i=g_i/2\pi v_F$.
Since the interaction is repulsive, $g$ is less than 1.
The renormalized velocity is given by
$v=v_F[(1+\tilde g_4)^2-(\tilde g_2)^2]^{1/2}$.
\end{multicols}
\vspace{-6mm}\noindent\underline{\hspace{87mm}}

We then introduce another set of bosonic fields
$\varphi_\pm(x)$:\cite{Affleck} 
\begin{equation}
\varphi_\pm(x)=
\frac{1}{\sqrt{8}}\left\{
\left(\frac{1}{\sqrt{g}}-\sqrt{g}\right)
[\varphi_R(x)\pm\varphi_L(-x)]
+\left(\frac{1}{\sqrt{g}}+\sqrt{g}\right)
[\varphi_R(-x)\pm\varphi_L(x)]
\right\}.
\label{varphi_pm}
\end{equation}
These fields satisfy
$[\varphi_+(x),\varphi_+(y)]=[\varphi_-(x),\varphi_-(y)]
  =-i\pi{\rm sgn}(x-y)$ and $[\varphi_+(x),\varphi_-(y)]=0$.
The advantage of using $\varphi_\pm$ is that we may separate the
Hamiltonian into two commuting parts, $H=H_F+H_B$, where
\begin{eqnarray}
H_F&=&
\frac{v}{4\pi}\int^\infty_{-\infty}dx
\left(\frac{d\varphi_-}{dx}\right)^2
+\frac{\lambda_F}{\pi}\sqrt{\frac{g}{2}}\frac{d\varphi_-(0)}{dx},
\label{H_F}
\\
H_B&=&
\frac{v}{4\pi}\int^\infty_{-\infty}dx
\left(\frac{d\varphi_+}{dx}\right)^2
+i\frac{\lambda_B}{\pi\alpha}\eta_L\eta_R
  \sin\left[\sqrt{2g}\varphi_+(0)+\theta\right].
\label{H_B}
\end{eqnarray}
The fermion field at $x=0$ may be written as
\begin{equation}
\psi(0)=
\frac{1}{\sqrt{2\pi\alpha}}
\exp\left[-\frac{i}{\sqrt{2g}}\varphi_-(0)\right] 
\left\{\eta_L\exp\left[-i\sqrt{\frac{g}{2}}\varphi_+(0)\right]
      +\eta_R\exp\left[i\sqrt{\frac{g}{2}}\varphi_+(0)\right]\right\}.
\label{psi(0)}
\end{equation}

\section{Local density of states at a scattering center}
\label{sec:ldos}

In this section we calculate the following correlation function:
\begin{equation}
D(t)\equiv
\langle g_\theta|
e^{iHt}\psi^\dagger(0)e^{-iHt}\psi(0)
|g_\theta\rangle,
\label{correlator}
\end{equation}
where $|g_\theta\rangle$ is a ground state of $H$.
The LDOS is given by
$\rho(\omega)=\int(d\omega/2\pi)e^{i\omega t}D(t)$.
In general we expect $D(t)\propto e^{-i\Delta t}t^{-\nu}$ for
$t\to\infty$.
Since $H$ has gapless excitations, we know that $\Delta$ must be zero.
Thus, we will not pay attention to $\Delta$ and concentrate only on
the exponent $\nu$ in the following discussion.

Since $H_F$ and $H_B$ commute, the correlation function is factorized
into two parts as $D(t)=\frac{1}{2\pi\alpha}D_F(t)D_B(t)$, where
\begin{mathletters}
\begin{eqnarray}
D_F(t)&=&
\langle{\rm F}|
e^{iH_Ft}e^{i\Phi_-}e^{-iH_Ft}e^{-i\Phi_-}|{\rm F}\rangle,
\label{D_F(t)}\\
D_B(t)&=&
\langle{\rm B}|e^{iH_Bt}
\left(\eta_Le^{i\Phi_+}+\eta_Re^{-i\Phi_+}\right)
e^{-iH_Bt}
\left(\eta_Le^{-i\Phi_+}+\eta_Re^{i\Phi_+}\right)
|{\rm B}\rangle,
\label{D_B(t)}
\end{eqnarray}
\end{mathletters}\noindent
Here $\Phi_-=\varphi_-(0)/\sqrt{2g}$,
$\Phi_+=\sqrt{g/2}\varphi_+(0)$, and $|{\rm F}\rangle$ ($|{\rm
B}\rangle$) is a ground states of $H_F$ ($H_B$).
The Hamiltonian $H_F$ is related to a free Hamiltonian by a unitary
transformation as
$UH_FU^\dagger=H_F^{(0)}+{\rm const}$, where
\begin{equation}
H_F^{(0)}=
\frac{v}{4\pi}\int^\infty_{-\infty}
dx\left(\frac{d\varphi_-}{dx}\right)^2
\label{UH_FU}
\end{equation}
and
\begin{equation}
U=
\exp\left[-i\frac{\lambda_F}{\pi v}\sqrt{\frac{g}{2}}
          \varphi_-(0)\right].
\end{equation}
This means $|F\rangle=U^\dagger|F_0\rangle$ with $|F_0\rangle$ being
the ground state of $H_F^{(0)}$.
We thus get
\begin{equation}
D_F(t)=
\langle F_0|
e^{iH_F^{(0)}t}e^{i\Phi_-}e^{-iH_F^{(0)}t}e^{-i\Phi_-}
|F_0\rangle
=\left(1+i\frac{vt}{\alpha}\right)^{-\frac{1}{2g}}
\sim t^{-\frac{1}{2g}}.
\label{D_F(t)-2}
\end{equation}
As pointed out in Ref.~\onlinecite{Oreg1}, the forward-scattering
potential does not affect the LDOS.

Next we rewrite Eq.~(\ref{D_B(t)}) as
\begin{equation}
D_B(t)=
\langle B|
e^{iH_Bt}\left(e^{i\Phi_+}-\eta_L\eta_Re^{-i\Phi_+}\right)
e^{-i\widetilde H_Bt}\left(e^{-i\Phi_+}+\eta_L\eta_Re^{i\Phi_+}\right)
|B\rangle,
\label{D_B(t)-2}
\end{equation}
\noindent\hspace{92mm}\underline{\hspace{87mm}}\vspace{-3mm}
\begin{multicols}{2}\noindent
where $\widetilde H_B\equiv\eta_LH_B\eta_L=H_B(\lambda_B\to-\lambda_B)$.
Note that this sign change of the cosine term is a direct consequence
of the anticommutation relation $\{\psi_L,\psi_R\}=0$.
At this point we may set $\eta_L\eta_R=-i$ because only the terms
involving even powers of $\eta_L\eta_R$ will contribute to $D_B(t)$
when Eq.~(\ref{D_B(t)-2}) is calculated perturbatively in powers of
$\lambda_B$.
We then shift
$\varphi_+(x)\to\varphi_+(x)+\frac{1}{\sqrt{2g}}(\frac{\pi}{2}-\theta)$
and obtain
\begin{eqnarray}
D_B(t)&=&
2\langle+|
e^{iH_+t}e^{i\Phi_+}e^{-iH_-t}e^{-i\Phi_+}
|+\rangle
\cr&&\cr&&
+2\cos\theta\langle+|
e^{iH_+t}e^{i\Phi_+}e^{-iH_-t}e^{i\Phi_+}
|+\rangle,
\label{D_B(t)-3}
\end{eqnarray}
where
\begin{equation}
H_\pm\equiv
\frac{v}{4\pi}\int^\infty_{-\infty}dx
\left(\frac{d\varphi_+}{dx}\right)^2
\pm\frac{\lambda_B}{\pi\alpha}\cos\left[\sqrt{2g}\varphi_+(0)\right]
\label{H_pm}
\end{equation}
and we have used the fact that the ground state of $H_+$, $|+\rangle$,
is invariant under $\varphi_+\to-\varphi_+$.
It is useful to transform Eq.~(\ref{D_B(t)-3}) further to the form
\begin{eqnarray}
D_B(t)&=&
2\langle+|e^{iH_+t}e^{-i\widetilde H_-t}|+\rangle\cr
&&
+2\cos\theta\langle+|e^{iH_+t}e^{-i\widetilde H_-t}e^{2i\Phi_+}
|+\rangle,
\label{D_B(t)-4}
\end{eqnarray}
where
\begin{eqnarray}
\widetilde H_-&=&
\frac{v}{4\pi}\int^\infty_{-\infty}dx
\left(\frac{d\varphi_+}{dx}-\pi\sqrt{2g}\delta(x)\right)^2\cr
&&
-\frac{\lambda}{\pi\alpha}\cos\left[\sqrt{2g}\varphi_+(0)\right].
\label{tildeH_-}
\end{eqnarray}

We first consider the case of $g=1/2$.
A crucial point in this case is that the cosine term
becomes $e^{i\varphi_+(0)}+e^{-i\varphi_+(0)}$.
Therefore, fermionizing the chiral boson $\varphi_+$ as
\begin{equation}
\frac{e^{i\varphi_+(x)}}{\sqrt{2\pi\alpha}}=\eta\psi_+(x),
\label{fermionization}
\end{equation}
we may transform Eq.~(\ref{H_pm}) to\cite{Matveev2,Guinea}
\begin{eqnarray}
H_\pm&=&
iv\int^\infty_{-\infty}dx\psi^\dagger_+(x)\frac{d}{dx}\psi_+(x)\cr
&&
\pm\frac{\lambda_B}{\sqrt{2\pi\alpha}}
\left[\eta\psi_+(0)+\psi^\dagger_+(0)\eta\right],
\label{H_pm-2}
\end{eqnarray}
where $\eta$ is a Majorana fermion, satisfying $\eta^2=1$.
This leads to a simple relation, $\eta H_+\eta=H_-$.
Note that $H_+$ is a quadratic Hamiltonian, which can be easily
diagonalized:\cite{Matveev2}
\begin{eqnarray}
H_+&=&
\int^\infty_{-\infty}dk
\left[\xi_ka_k^\dagger a_k
      +\frac{\lambda_B}{2\pi\sqrt\alpha}
            \left(\eta a_k+a_k^\dagger\eta\right)
\right]\cr
&=&
\int^\infty_0dk\xi_k\left(c^\dagger_kc_k+d^\dagger_kd_k\right)
+{\rm const},
\end{eqnarray}
where $\xi_k\equiv vk$ and
$\psi_+(x)=\int(dk/\sqrt{2\pi})e^{-ikx}a_k$.
For later convenience we write the transformation rule
here:\cite{Matveev2}
\end{multicols}
\vspace{-6mm}\noindent\underline{\hspace{87mm}}
\begin{mathletters}
\begin{eqnarray}
a_k&=&
\frac{1}{\sqrt2}c_k+\frac{\xi_k}{\sqrt{2(\xi^2_k+\Gamma^2)}}d_k
+\frac{\Gamma}{\sqrt{2}\pi}{\rm P}\!\!\int^\infty_0dq
   \frac{1}{\sqrt{\xi^2_q+\Gamma^2}}
   \left(\frac{d_q}{q-k}-\frac{d^\dagger_q}{q+k}\right),
\label{a_k}\\
a_{-k}&=&
\frac{1}{\sqrt2}c^\dagger_k
-\frac{\xi_k}{\sqrt{2(\xi^2_k+\Gamma^2)}}d^\dagger_k
+\frac{\Gamma}{\sqrt{2}\pi}{\rm P}\!\!\int^\infty_0dq
   \frac{1}{\sqrt{\xi^2_q+\Gamma^2}}
   \left(\frac{d_q}{q+k}-\frac{d^\dagger_q}{q-k}\right),
\label{a_-k}\\
\eta&=&
\frac{\lambda_B}{\pi}\sqrt{\frac{2}{\alpha}}\,
{\rm P}\!\!\int^\infty_0dq\frac{1}{\sqrt{\xi^2_q+\Gamma^2}}
\left(d_q+d^\dagger_q\right),
\label{eta}
\end{eqnarray}
\end{mathletters}\noindent
where $k>0$, $\Gamma\equiv\lambda^2_B/(\pi\alpha v)$, and
$c_k$ and $d_k$ satisfy the ordinary anticommutation relation.
The ground state $|+\rangle$ is the vacuum of $c_k$ and $d_k$.

Using Eq.~(\ref{fermionization}), we rewrite Eq.~(\ref{D_B(t)-4})
in a fermionic form,
\begin{equation}
D_B(t)=
2\langle+|e^{iH_+t}\eta e^{-i\widetilde H_+t}\eta|+\rangle
+2\sqrt{2\pi\alpha}\cos\theta
\langle+|e^{iH_+t}\eta e^{-i\widetilde H_+t}\psi_+(0)|+\rangle,
\label{D_B(t)-5}
\end{equation}
where
\begin{equation}
\widetilde H_+
=H_+ + \pi v:\!\psi_+^\dagger(0)\psi_+(0)\!: +{\rm const}.
\label{tildeH_+}
\end{equation}
From Eqs.~(\ref{a_k}) and (\ref{a_-k}), the second term becomes
\begin{equation}
\pi v:\!\psi^\dagger_+(0)\psi_+(0)\!:\,\,
=\frac{v}{2}\int^\infty_0dk\int^\infty_0dp
\frac{\xi_p}{\sqrt{\xi^2_p+\Gamma^2}}
\left(c_k+c^\dagger_k\right)\left(d_p-d^\dagger_p\right),
\label{psipsi}
\end{equation}
\noindent\hspace{92mm}\underline{\hspace{87mm}}\vspace{-3mm}
\begin{multicols}{2}\noindent
which is an irrelevant operator with scaling dimension 2.
To find the long-time behavior of $D_B(t)$, we can thus treat
Eq.~(\ref{psipsi}) as a small perturbation.
The lowest-order calculation then gives, for $\Gamma t\gg1$,
\begin{equation}
D_B(t)=
-\frac{4i}{\pi\Gamma t}
+\sqrt{2\pi\alpha}\cos\theta\frac{\lambda_B}{\pi v}
 \frac{\ln(vt/\alpha)}{\Gamma^2t^2}.
\label{D_B(t)-6}
\end{equation}
Note that the $1/t$-dependence of the first term comes from the
correlator $\langle+|\eta(t)\eta(0)|+\rangle$, which also appeared in
the two-channel Kondo problem.\cite{Emery}
Combining Eqs.~(\ref{D_F(t)-2}) and (\ref{D_B(t)-6}), we get
$D(t)=-2/(\pi^2v\Gamma t^2)$ for $\Gamma t\gg1$, which implies
\begin{equation}
\rho(\omega)=\frac{2\omega}{\pi^2v\Gamma}
\end{equation}
for $\omega\ll\Gamma$.
This is consistent with Eq.~(\ref{rho(omega)}).
We see that the single scatterer at $x=0$ indeed depletes the 
low-energy excitations around it.

For $g\ne1/2$ ($0<g<1$) we take a different approach.
We assume from the outset that the phase field $\varphi_+$ is
pinned at $x=0$ by the cosine potential in $H_+$ (\ref{H_pm}), as in
Refs.~\onlinecite{Gogolin,Prokofev}, and \onlinecite{Affleck}.
We thus replace the cosine by a term which is easier to deal with.
A convenient choice is
\begin{equation}
H_M=
\frac{v}{2}\int^\infty_{-\infty}dx
 \left[\frac{1}{g}\left(\frac{d\phi}{dx}\right)^2+g\Pi^2\right]
+\frac{M}{2}[\phi(0)]^2,
\label{H_M}
\end{equation}
where $M$ should be a characteristic energy scale at which the cosine
term becomes of the order of the band width ($M=\Gamma$ for $g=1/2$).
It immediately follows from the scaling equation
$d\lambda_B/dl=(1-g)\lambda_B$ that
\begin{equation}
M\propto\frac{v}{\alpha}\left(\frac{\lambda_B}{v}\right)^{\frac{1}{1-g}}.
\label{M}
\end{equation}
Since $H_M$ is a quadratic Hamiltonian, it is easily diagonalized as
$H_M=\int dk\xi_k
     \left(\alpha^\dagger_k\alpha_k+\beta^\dagger_k\beta_k\right)$
with
\end{multicols}
\begin{equation}
\phi=
\int^\infty_0dk\sqrt{\frac{g}{2\pi k}}
\left[
\sin(kx)\left(\alpha_k+\alpha_k^\dagger\right)
+\cos(k|x|-\delta_k)\left(\beta_k+\beta_k^\dagger\right)
\right]
\label{H_M-2}
\end{equation}
and $\Pi=(1/gv)\partial\Pi/\partial t$, where $\alpha_k$ and $\beta_k$
satisfy the ordinary commutation relations of bosons.
The phase shift is given by $\delta_k=\tan^{-1}(gM/2vk)$.
Note that $\delta_k\to\pi/2$ as $k\to0$.

Let us denote the ground state of $H_M$ by $|0_M\rangle$.
We then find
\begin{equation}
\langle0_M|
\partial_x\varphi_+(0,t)\,\partial_x\varphi_+(0,0)
|0_M\rangle
=
2\pi g\langle0_M|\Pi(0,t)\Pi(0,0)|0_M\rangle
=
\frac{24}{g^2M^2v^2t^4}
\label{t^-4}
\end{equation}
\noindent\hspace{92mm}\underline{\hspace{87mm}}\vspace{-3mm}
\begin{multicols}{2}\noindent
for $Mt\gg1$, implying that $\partial_x\varphi_+(0)$ is an irrelevant
operator with dimension 2.
This is consistent with the observation made in Eq.~(\ref{psipsi}).
In fact, this is an expected result because $\varphi_+$ is pinned at
$x=0$. 
We may thus use $H_-$ instead of $\widetilde H_-$ to obtain the
long-time asymptotic behavior of $D_B(t)$ in Eq.~(\ref{D_B(t)-4}).
It is also important to note that $e^{i\Phi_+}$ is not fluctuating too 
much and can be regarded essentially as a constant because
$\varphi_+(0)$ is pinned. 
In fact, we find
\begin{eqnarray}
\langle0_M|e^{i\Phi_+}|0_M\rangle&=&
\langle0_M|\exp\left[i\sqrt{2\pi/g}\phi(0)\right]|0_M\rangle\cr
&=&\sqrt{\frac{e^\gamma g\alpha M}{2v}}
\label{e^iPhi_+}
\end{eqnarray}
for $\alpha M\ll v$, where $\gamma=0.577\ldots$ is Euler's constant.
Note that, at $g=1/2$, we get
$\langle+|e^{i\Phi_+}|+\rangle=-(\lambda_B/\pi v)\ln(v/\alpha\Gamma)$, 
which is consistent with Eqs.~(\ref{M}) and (\ref{e^iPhi_+}).
Hence, from Eq.~(\ref{D_B(t)-4}), we get
\begin{eqnarray}
D_B(t)&\propto&
\langle0_M|e^{iH_+t}Ve^{-iH_+t}V^\dagger|0_M\rangle\cr
&\approx&
\langle0_M|e^{iH_Mt}Ve^{-iH_Mt}V^\dagger|0_M\rangle,
\label{VV}
\end{eqnarray}
where $V$ is a unitary operator which shifts
$\phi(x)\to\phi(x)+\frac{\sqrt\pi}{2}$.
The rhs of Eq.~(\ref{VV}) is known to decay as
$\sim t^{-1/2g}$.\cite{Fabrizio}
This result can be easily obtained using the following representation
for $V$:
\begin{eqnarray}
V&=&\exp\left[i\frac{\sqrt\pi}{2}\int dx\Pi(x)\right]\cr
&=&
\exp\left[-\int^\infty_0dk\frac{e^{-\alpha k}}{\sqrt{2gk}}
            \sin\delta_k\left(\beta^\dagger_k-\beta_k\right)\right].
\label{V}
\end{eqnarray}
We therefore conclude $D(t)\propto t^{-1/g}$, from which
Eq.~(\ref{rho(omega)}) follows.
We emphasize that the above calculation should give the exact value of 
the exponent, although the amplitude may not be correct.

\section{Orthogonality catastrophe}
\label{sec:ortho}

In this section we discuss the orthogonality catastrophe for the
special case of $g=1/2$.
We calculate the overlap integral
$|\langle{\rm p}|{\rm s}\rangle|^2=
|\langle F_0|F\rangle|^2\times|\langle0|+\rangle|^2$, where $|0\rangle$
is the ground state of the Hamiltonian $H_0\equiv H_+|_{\lambda_B=0}$.
It is almost trivial to find $\gamma_F$ in
Eq.~(\ref{-gamma_F-gamma_B}) because
$\langle F_0|F\rangle=\langle F_0|U^\dagger|F_0\rangle$.
We get\cite{Ogawa,Lee} 
\begin{equation}
\gamma_F=2g\left(\frac{\lambda_F}{2\pi v}\right)^2.
\end{equation}
Hence our problem is reduced to calculate the overlap
$\langle0|+\rangle$.
In the fermion language, $H_0$ is
\begin{equation}
H_0=iv\int^\infty_{-\infty}dx\psi^\dagger_+(x)\frac{d}{dx}\psi_+(x),
\label{H_0}
\end{equation}
and $|0\rangle$ is the filled Fermi sea.
Then the ground state of $H_+$ can be written as
\begin{equation}
|+\rangle={\rm T}\exp\left[-i\int^0_{-\infty}
e^{\epsilon t}H'(t)dt\right]|0\rangle,
\end{equation}
where $\epsilon$ is positive infinitesimal and
$H'(t)=e^{iH_0t}(H_+-H_0)e^{-iH_0t}$.
Using the linked-cluster theorem, we can write the overlap
integral as
\begin{equation}
\langle0|+\rangle
=\exp[G_c(0,-\infty)],
\label{linked-cluster}
\end{equation}
where $G_c(0,-\infty)$ is a sum of connected ring diagrams,
\end{multicols}
\vspace{-6mm}\noindent\underline{\hspace{87mm}}
\begin{eqnarray}
G_c(0,-\infty)&=&
-\sum^\infty_{n=1}\frac{\lambda^{2n}}{2n}
 \int^0_{-\infty}\!\!dt_1\cdots\int^0_{-\infty}\!\!dt_{2n}
 s_0(t_1-t_2)g_0(t_2-t_3)\cdots s_0(t_{2n-1}-t_{2n})g_0(t_{2n}-t_1)
 \exp\!\left(\sum^{2n}_{i=1}\epsilon t_i\right).
\label{G_c}
\end{eqnarray}
Here $\lambda\equiv\lambda_B/\sqrt{2\pi\alpha}$ and the propagators
$s_0(t)$ and $g_0(t)$ are given by
\begin{mathletters}
\begin{eqnarray}
s_0(t)&=&\langle0|{\rm T}\eta(t)\eta(0)|0\rangle={\rm sgn}(t),\\
g_0(t)&=&
\langle0|{\rm T}[\psi_+(x=0,t)-\psi^\dagger_+(0,t)]
                [\psi_+(0,0)-\psi^\dagger_+(0,0)]|0\rangle
=\frac{i}{\pi v[t-i\varepsilon{\rm sgn}(t)]},
\label{g_0}
\end{eqnarray}
\end{mathletters}\noindent
where $\varepsilon$ is positive infinitesimal.
Differentiating Eq.~(\ref{G_c}) with respect to $\lambda$, we
obtain
\begin{equation}
G_c(0,-\infty)=
-\frac{v}{4}\int^\Gamma_0d\Gamma
\int^0_{-\infty}dt_1\int^0_{-\infty}dt_2 e^{\epsilon(t_1+t_2)}
s_0(t_1-t_2)g(t_2,t_1),
\label{G_c-2}
\end{equation}
where $g(t_1,t_2)$ is a solution of a Dyson equation, 
\begin{equation}
g(t_1,t_2)=g_0(t_1-t_2)
-\frac{\Gamma}{2\pi i}{\rm P}\!\!
\int^0_{-\infty}\!\!dt_3\int^0_{-\infty}\!\!dt_4
\frac{e^{\epsilon(t_3+t_4)}}{t_1-t_3}{\rm sgn}(t_3-t_4)g(t_4,t_2).
\label{Dyson1}
\end{equation}
Since Eq.~(\ref{Dyson1}) contains double integral, working in real
time is not so convenient as it is in the Fermi-liquid
case.\cite{Nozieres} 
On the other hand, the Fourier transform of Eq.~(\ref{Dyson1})
contains only a single integral:
\begin{equation}
\tilde g(\omega,t_2)=
-\frac{e^{i\omega t_2}}{v}{\rm sgn}(\omega)
+\frac{i\Gamma}{|\omega|}\int^\infty_{-\infty}\frac{d\nu}{2\pi i}
\tilde g(\nu,t_2)\left[\frac{1}{\nu-\omega+2i\epsilon}
                      -\frac{1}{2(\nu+i\epsilon)}\right].
\label{Dyson2}
\end{equation}
This equation can be solved in the limit $\epsilon\to0$ in the
standard way.\cite{Musk}

We first introduce functions $\tilde g_\pm$ by
\begin{equation}
\tilde g_\pm(\omega)=
\int^\infty_{-\infty}\frac{d\nu}{2\pi i}
\tilde g(\nu,t_2)\left[\frac{1}{\nu-\omega\mp2i\epsilon}
                      -\frac{1}{2(\nu+i\epsilon)}\right].
\end{equation}
We can then express Eq.~(\ref{Dyson2}) as
\begin{equation}
\tilde g_+(\omega)
-\left(1-i\frac{\Gamma}{|\omega|}\right)\tilde g_-(\omega)=
-\frac{e^{i\omega t_2}}{v}{\rm sgn}(\omega).
\label{Dyson3}
\end{equation}
A solution of this equation with correct analytic properties is
\begin{equation}
\tilde g_\pm(\omega)=
-\frac{1}{v}\int^\infty_{-\infty}\!\frac{d\nu}{2\pi i}
 \frac{e^{i\nu t_2}{\rm sgn}(\nu)}{\nu-\omega\mp i\delta}
 \frac{X_\pm(\omega)}{X_+(\nu)},
\label{solution}
\end{equation}
where $\delta$ is positive infinitesimal and
\begin{equation}
X_\pm(\omega)=
\exp\!\left[\int^\infty_{-\infty}\!\frac{d\nu}{2\pi i}
            \frac{\ln\left(1-i\frac{\Gamma}{|\nu|}\right)}
                 {\nu-\omega\mp i\delta}\right].
\label{X_pm}
\end{equation}
With this solution Eq.~(\ref{G_c-2}) becomes
\begin{eqnarray}
G_c(0,-\infty)&=&
\frac{1}{8\pi^2i}\int^\Gamma_0\!\!d\Gamma
\int^0_{-\infty}\!\!dt_1\!\int^{t_1}_{-\infty}\!\!dt_2\!
\int^\infty_{-\infty}\!\!d\omega\!\int^\infty_{-\infty}\!\!d\nu\,
e^{-i\omega t_1+i\nu t_2+\epsilon(t_1+t_2)}
\frac{{\rm sgn}(\nu)}{\nu-\omega+i\delta}
\frac{X_-(\omega)}{X_+(\nu)}\cr
&=&
\frac{1}{8\pi^2}\int^\Gamma_0\!\!d\Gamma
\int^0_{-\infty}\!\!d\tau\!\int^\infty_{-\infty}\!\!d\nu\,
e^{i(\nu-i\epsilon)\tau}\frac{{\rm sgn}(\nu)}{X_+(\nu)}
\int^\infty_{-\infty}\!\!d\omega
\frac{X_-(\omega)}{(\omega-\nu+2i\epsilon)(\nu-\omega+i\delta)},
\label{G_c-3}
\end{eqnarray}
where $\tau=t_2-t_1$ and we have integrated over $(t_1+t_2)/2$.
As pointed out by Hamann,\cite{Hamann} in the next step in which we
perform the $\omega$ integral, it is important to keep $\epsilon$
finite while taking the limit $\delta\to+0$:
\begin{eqnarray*}
G_c(0,-\infty)&=&
-\frac{i}{4\pi}\int^\Gamma_0\!\!d\Gamma\!
\int^0_{-\infty}\!\!d\tau\!\int^\infty_{-\infty}\!\!d\nu
e^{i(\nu-i\epsilon)\tau}\frac{{\rm sgn}(\nu)}{2i\epsilon}
\frac{X_-(\nu-2i\epsilon)}{X_+(\nu)}\cr
&=&
-\frac{1}{8\pi\epsilon}\int^\Gamma_0\!\!d\Gamma\!
 \int^0_{-\infty}\!\!d\tau\!\int^\infty_{-\infty}\!\!d\nu
 \frac{\nu e^{i(\nu-i\epsilon)\tau}}{|\nu|-i\Gamma}
+\frac{1}{8\pi^2}\int^\Gamma_0\!\!d\Gamma\!
 \int^0_{-\infty}\!\!d\tau\!\int^\infty_{-\infty}\!\!d\nu_1\!
 \int^\infty_{-\infty}\!\!d\nu_2
 \frac{\nu_1e^{i(\nu_1-i\epsilon)\tau}}{|\nu_1|-i\Gamma}
 \frac{\ln\left(1-i\frac{\Gamma}{|\nu_2|}\right)}
      {(\nu_2-\nu_1+i\delta)^2}.
\end{eqnarray*}
After replacing $\nu/(|\nu|-i\Gamma)$ by
$\nu[(|\nu|-i\Gamma)^{-1}-(\nu-i\Gamma)^{-1}]$ and
$\ln\bigl(1-i\frac{\Gamma}{|\nu|}\bigr)$ by
$\ln\bigl[\bigl(1-i\frac{\Gamma}{|\nu|}\bigr)
          /\left(1+i\frac{\Gamma}{\nu}\right)\bigr]$,
we integrate over $\tau$ to obtain
\begin{eqnarray}
G_c(0,-\infty)&=&
\frac{1}{4\pi\epsilon i}\int^\Gamma_0\!\!d\Gamma\!
\int^0_{-\infty}\!\!d\nu\frac{\nu}{\nu^2+\Gamma^2}
-\frac{1}{2\pi^2}\int^\Gamma_0\!\!d\Gamma\!
\int^\infty_0\!\!d\nu_1\!\int^\infty_0\!\!d\nu_2
\frac{1}{(\nu_1+\nu_2)^2}
\frac{\nu_1}{\nu_1^2+\Gamma^2}
\tan^{-1}\left(\frac{\Gamma}{\nu_2}\right)\cr
&=&
\frac{i}{2\epsilon}\frac{\Gamma}{2\pi}
\left[\ln\!\left(\frac{\Lambda}{\Gamma}\right)+1\right]
-\frac{1}{16}\ln\!\left(\frac{\Gamma}{E_L}\right),
\label{G_c-5}
\end{eqnarray}
where we have introduced the high-energy cutoff $\Lambda\sim v/\alpha$ 
and the low-energy cutoff $E_L\sim v/L$.
From Eqs.\ (\ref{linked-cluster}) and (\ref{G_c-5}) we get
$\gamma_B=1/8$ in agreement with the previous
studies.\cite{Gogolin,Prokofev,KMG,Affleck,Qin} 
Note that the quantity
$E_0\equiv-(\Gamma/2\pi)[\ln(\Lambda/\Gamma)+1]$ appearing in the
first term is equal to the difference between the ground state
energies of $H_+$ and $H_0$.\cite{Matveev2}

Since $\delta(E)\equiv\tan^{-1}(\Gamma/E)$ in Eq.~(\ref{G_c-5}) is the
phase shift for fictitious chiral fermions due to the coupling
$\lambda_B$ in Eq.~(\ref{H_pm-2}), the above calculation implies that
$\gamma_B=\frac{1}{2}[\delta(0)/\pi]^2$, in contrast to the
Fermi-liquid result\cite{Anderson,Nozieres}
$\gamma_{\rm Fermi}=[\delta(0)/\pi]^2$.
The extra factor $1/2$ in our result can be traced back to
the peculiar form of the scattering term in Eq.~(\ref{H_pm-2}).
Only the combination $\psi_+-\psi^\dagger_+$ interacts with $\eta$, and
the other combination $\psi_++\psi^\dagger_+$ is decoupled.
Hence only {\it half} of the degrees of freedom have the phase shift
($\delta(0)=\pi/2$), giving the factor $1/2$.

As pointed out by Matveev,\cite{Matveev2} the Hamiltonian (\ref{H_pm-2})
is equivalent to the effective Hamiltonian of the two-channel Kondo
model in the Toulouse limit,\cite{Emery} where the Majorana fermion
$\eta$ corresponds to the $xy$-component of the impurity spin.
Thus our calculation also applies to the orthogonality catastrophe in 
the two-channel Kondo problem in which $J_\perp$ is turned on and off
while $J_z$ kept constant.

\section{Core-hole Green's function}
\label{sec:core-hole}

Next we calculate the core-hole Green's function,
\begin{equation}
G(t)=\langle0|e^{iH_0t}e^{-iH_+t}|0\rangle
\label{G(t)}
\end{equation}
for $g=1/2$.
Using the linked-cluster theorem again, we get $G(t)=\exp[G_c(t,0)]$,
where $G_c(t,0)$ is
\begin{equation}
G_c(t,0)=
-\sum^\infty_{n=1}\frac{\lambda^{2n}}{2n}
 \int^t_0\!\!dt_1\cdots\int^t_0\!\!dt_{2n}
 s_0(t_1-t_2)g_0(t_2-t_3)\cdots s_0(t_{2n-1}-t_{2n})g_0(t_{2n}-t_1).
\label{G_c-6}
\end{equation}
\noindent\hspace{92mm}\underline{\hspace{87mm}}\vspace{-3mm}
\begin{multicols}{2}\noindent
This time we differentiate Eq.~(\ref{G_c-6}) with respect to $t$ to
get
\begin{equation}
-\frac{d}{dt}G_c(t,0)=\lambda^2\int^t_0\!\!dt_1g_0(t-t_1)s(t_1),
\label{d/dt}
\end{equation}
where $s(t_1)$ is defined for $0\le t_1\le t$ and is a solution of a
Dyson equation, 
\begin{equation}
s(t_1)=-1
-\frac{\Gamma}{2\pi i}{\rm P}\!\!\int^t_0\!\!dt_3\!
\int^t_0\!\!dt_4\frac{{\rm sgn}(t_1-t_3)}{t_3-t_4}s(t_4).
\label{Dyson-2}
\end{equation}
From this equation we can easily show that
$s(t_1)=s(t-t_1)$ and $s(+0)=-1$.
Thus Eq.~(\ref{d/dt}) becomes
\begin{equation}
-\frac{d}{dt}G_c(t,0)=
\frac{\Gamma}{4}
-\frac{\Gamma}{2\pi i}\int^t_0\!\!dt_1\frac{s(t_1)}{t_1}.
\label{d/dt-2}
\end{equation}
Here the first term comes from the real part of $g_0$ in
Eq.~(\ref{g_0}).

For short times $\Gamma t\ll1$, we can solve Eq.~(\ref{Dyson-2})
perturbatively.
Up to order $(\Gamma t)^2$ we obtain
\begin{equation}
G_c(t,0)=
i\frac{\Gamma t}{2\pi}\left[\ln\!\left(\frac{t}{t_c}\right)-1\right]
-\frac{1}{4}\Gamma t+\frac{1}{24}(\Gamma t)^2,
\label{short-time}
\end{equation}
where $t_c$ is a short-time cutoff $\sim1/\Lambda$.
This expansion, however, starts to fail around $\Gamma t\sim1$.
From the analysis in Sec.~\ref{sec:ortho}, for $\Gamma t\gg1$
we expect $G_c(t,0)$ to approach
$-iE_0t-\frac{1}{8}\ln(\Gamma t)$.\cite{Gogolin,Prokofev,KMG,Affleck}

The crossover from the short-time to the long-time regimes can be seen 
most conveniently by solving Eq.~(\ref{Dyson-2}) numerically and
putting the solution into Eq.~(\ref{d/dt-2}).
Note that the integral in Eq.~(\ref{d/dt-2}) is well-defined because
${\rm Im}s(t_1)\sim t_1|\ln t_1|$ for $t_1\to0$.
Figure 1 shows the $t$-dependence of the real part of $(d/dt)G_c(t,0)$ 
computed in this way.
It clearly exhibits the crossover at $\Gamma t\sim1$ from the
short-time behavior, Eq.~(\ref{d/dt-2}), to the long-time asymptote,
${\rm Re}[dG_c(t,0)/dt]=-1/8t$.
\begin{figure}
\narrowtext
\begin{center}\leavevmode\epsfysize=65mm \epsfbox{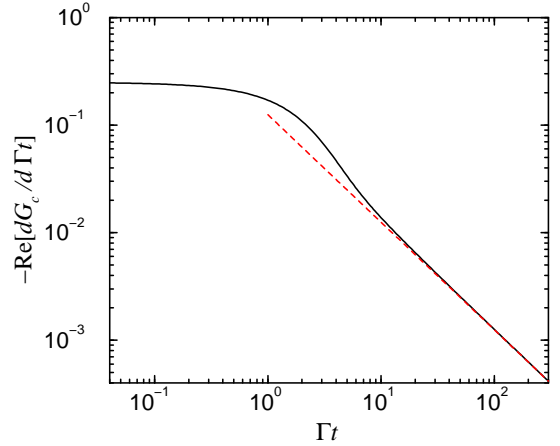}\end{center}
\caption{Time evolution of the core-hole Green's function.  There is a
clear crossover at $\Gamma t\sim1$.  The dashed line represents
${\rm Re}[dG_c/d\Gamma t]=-1/(8\Gamma t)$.} 
\end{figure}

\section{Fermi-edge singularity}
\label{sec:fermi-edge}

In this section we briefly discuss the Fermi-edge singularity for
$g<1$ to show that the exponents can be easily obtained from the
analysis of Secs.~\ref{sec:ortho} and \ref{sec:core-hole}. 
Here we are concerned with the correlation function
\begin{equation}
I(t)=
\langle g_0|e^{i(H_F^{(0)}+H_0)t}\psi(0)
            e^{-i(H_F+H_B)t}\psi^\dagger(0)|g_0\rangle,
\label{I(t)}
\end{equation}
where $|g_0\rangle\equiv|F_0\rangle\otimes|0\rangle$.
Following the same path as in Sec.~\ref{sec:ldos}, we write the
correlator as $I(t)=\frac{1}{2\pi\alpha}I_F(t)I_B(t)$,
where\cite{Ogawa,Lee} 
\begin{eqnarray}
I_F(t)&=&
\langle F_0|
e^{iH_F^{(0)}t}e^{-i\Phi_-}Ue^{-iH_F^{(0)}t}U^\dagger e^{i\Phi_-}
|F_0\rangle\cr
&\sim& t^{-\nu_F}
\label{I_F(t)}
\end{eqnarray}
with
$\nu_F=\left(\frac{1}{\sqrt{2g}}
          +\frac{\lambda_F}{2\pi v}\sqrt{2g}\right)^2$
and
\begin{eqnarray}
I_B(t)&=&
2\langle 0|e^{iH_0t}e^{-i\widetilde H_-t}|0\rangle\cr
&&
+2\cos\theta
\langle 0|e^{iH_0t}e^{-i\widetilde H_-t}e^{2i\Phi_+}|0\rangle.
\label{I_B(t)}
\end{eqnarray}
We expect that $I_B(t)$ should decay as $I_B(t)\propto t^{-\nu_B}$ in
the long-time limit.
We now notice that the first term in Eq.~(\ref{I_B(t)}) is similar to
the core-hole Green's function discussed in Sec.~\ref{sec:core-hole}.
As we saw in Fig.~1, it should decay as $\sim t^{-\tilde\gamma}$ with 
$\tilde\gamma$ being the exponent of the orthogonality catastrophe
between $|0\rangle$ and the ground state of $\widetilde H_-$:
$|\langle0|-\rangle|^2\propto L^{-\tilde\gamma}$.
The latter state has a finite overlap with the ground state of $H_-$,
because $\partial_x\varphi(0)$ $[\propto(\widetilde H_- - H_-)]$ is an
irrelevant operator around the fixed point of $H_-$.
This means $\tilde\gamma=\gamma_B=1/8$.
Since the second term in Eq.~(\ref{I_B(t)}) contains extra factor,
$e^{2i\Phi_+}$, at least it is not larger than the first term.
Hence we conclude $\nu_B=1/8$, in agreement with
Refs.~\onlinecite{Prokofev} and \onlinecite{Affleck}. 
The fact that $\nu_B$ equals $\gamma_B$ is a direct consequence of the
pinning of $\varphi_+$ at $x=0$.
Therefore the insertion of $\varphi_+$ part of the fermion field,
$e^{i\Phi_+}$, does not change the exponent.
On the other hand, $\nu_F$ is not equal to $\gamma_F$ because the
forward scattering potential is a marginal operator.

\section{Conclusion}
\label{sec:conclusion}

In this paper we have studied the low-energy behavior of the LDOS at
the location of a scattering center and the orthogonality catastrophe
due to a sudden local perturbation.
The characteristic, anomalous low-energy (long-time) properties were
obtained by exact calculations for $g=1/2$ by mapping the bosonized
Hamiltonian back to a fermionic quadratic Hamiltonian.
This method has allowed us to describe the crossover from the
weak-coupling (short-time) to the strong-coupling (long-time) regimes.
The exact results obtained for $g=1/2$ agree with the previous studies 
based on the assumption that the phase fields are completely pinned at
the impurity site in the low-energy limit.
The agreement implies that, to describe the low-energy physics, it is
sufficient to use an effective model which incorporates the perfect
reflection by the local potential.
We conclude that $\gamma_B=1/8$ and
$\rho(\omega)\propto\omega^{\frac{1}{g}-1}$ for $g<1$.
It seems that the mapping to a Coulomb gas problem used in
Refs.~\onlinecite{Oreg1} and \onlinecite{Oreg2} makes it difficult to
capture the Majorana fermions which have played an essential role in
this paper. 

After completion of this work the author became aware that Fabrizio
and Gogolin\cite{comment} obtained a similar result on the
low-energy behavior of the LDOS, Eq.~(\ref{rho(omega)}).

\acknowledgements

The author would like to thank N.~Kawakami, N.~Nagaosa, and
V.~Ponomarenko for helpful discussions.
The numerical computation was supported by the Yukawa Institute for
Theoretical Physics and also done in part on VPP500 at the
Institute for Solid State Physics, University of Tokyo.

\end{multicols}

\end{document}